\documentclass[twocolumn,showpacs,amsmath,amstex,amssymb,mathfonts,prl]{revtex4}
\usepackage{graphicx,bm,units}

\begin{document}

\newcommand{\vk}{{\bf k}}
\def\ns{^{\vphantom{*}}}
\def\ket#1{{|  #1 \rangle}}
\def\bra#1{{\langle #1|}}
\def\braket#1#2{{\langle #1  |   #2 \rangle}}
\def\expect#1#2#3{{\langle #1 |   #2  |  #3 \rangle}}
\def\cH{{\cal H}}
\def\half{\frac{1}{2}}
\def\sut{\textsf{SU}(2)}
\def\suto{\textsf{SU}(2)\ns_1}
\def\kF{\ket{\,{\rm F}\,}}
\newcommand{\ack}[1]{[{\bf Pfft!: {#1}}]}

\title{Torsional Response and Dissipationless Viscosity in Topological Insulators}

\author{Taylor L. Hughes}\author{Robert G. Leigh}\author{Eduardo Fradkin}
\address{Department of Physics, University of Illinois, 1110 West Green St, Urbana IL 61801}

\begin{abstract} 
We consider the visco-elastic response of the electronic degrees of freedom in 2D and 3D topological insulators (TI). Our primary focus is on the 2D Chern insulator which exhibits a bulk dissipationless viscosity analogous to the quantum Hall viscosity predicted in integer and fractional quantum Hall states. We show that the  dissipationless viscosity is the response of  a TI  to torsional deformations of the underlying lattice geometry. The visco-elastic response also indicates that crystal dislocations in Chern insulators will carry momentum density. We briefly discuss generalizations to 3D which imply that time-reversal invariant TI's will exhibit a quantum Hall viscosity on their surfaces. 
\end{abstract}


\date{\today}

\maketitle
A striking feature of a topological insulator (TI)  is its topological response. The paradigmatic example is the time-reversal breaking integer quantum Hall effect (IQHE) in 2D which exhibits a Hall conductance that is an integer multiple of $e^2/h$\cite{prangeBook}. More recently it was shown that there exist related states in 3D which are time-reversal invariant\cite{hasan2010,qi2010} and exhibit a topological magneto-electric effect\cite{qi2008B} (TME). While the electro-magnetic response of topological insulators is the most well-known, in this Letter we consider the visco-elastic response of the electronic degrees of freedom in TI's. Namely, we want to consider the stress response
\begin{equation}
\langle T^{ij}\rangle= \Lambda^{ijk\ell}u_{k\ell}+\eta^{ijk\ell}\dot{u}_{k\ell}
\end{equation}\noindent where $T^{ij}$ is the stress-tensor, $\Lambda,\eta$ are the elasticity and viscosity tensors respectively and $u_{ij}$ is the strain tensor.  Here we show there is a dissipationless viscosity response in the topological Chern insulator\cite{Haldane1988} (CI) state analogous to that found in the IQHE and fractional QHE states\cite{avron1995,read2009,tokatly2009,haldane2009A,read2010}. While viscosities are normally associated with frictional dissipation, this viscosity, present only when time-reversal symmetry is broken,  implies a force perpendicular to the fluid motion similar to the Lorentz force.  

In a condensed matter system the electronic stress response can be calculated by coupling the electronic Hamiltonian to perturbations of the background lattice geometry. The \emph{topological} responses due to geometric \emph{curvature} have been studied in Refs. \cite{ryu2010,qi2010A} in the language of quantum field theory anomalies.  Alternatively, we consider the response of topological insulators to an external \emph{torsion} field. A heuristic understanding of the difference between curvature and torsion is that when an object traverses a small loop in real space it is rotated if there is non-zero curvature, and translated if there is non-zero torsion. A familiar manifestation of torsion is a crystal dislocation. These line-defects are singular sources of torsion, analogous to a localized magnetic flux line. For example,  while dragging an electron  around a magnetic flux line its wavefunction is multiplied by a $U(1)$ phase, while for a dislocation line it is multiplied by a translation operator along the Burgers vector. The visco-elastic response is derived as a linear response to  torsional perturbations of the underlying material geometry.  We find that this response also leads to momentum density localized on crystal dislocations and we also briefly mention the visco-elastic response  of 3D time-reversal invariant topological insulators (3DTI).

To understand the torsion response we will often draw comparisons to the well-known electromagnetic responses of topological insulators which we briefly review now.  The responses of the CI and 3DTI to external electromagnetic fields  are encapsulated in topological effective actions \emph{i.e.} free-energy functionals derived from calculating a partition function in the presence of external fields. The QHE and TME are encoded in the effective actions
\begin{eqnarray}
S_{eff}[A_{\mu}]&=&\frac{n e^2}{2h}\int d^3 x \epsilon^{\mu\nu\rho}A_{\mu}\partial_{\nu}A_{\rho}\label{eq:CS_A}\\
S_{eff}[A_{\mu}]&=&\frac{e^2}{16 h}\int d^4 x\;\theta \;\epsilon^{\mu\nu\sigma\tau}F_{\mu\nu}F_{\sigma\tau}\label{eq:FF_A}
\end{eqnarray}\noindent  respectively which are derived from the responses of the topological insulators to an external field $A_\mu$ (note that  $n$ is an integer). The nominal current response is $\langle j^{\mu}\rangle=\delta S_{eff}/\delta A_{\mu}$ which gives the QHE and TME when acting on Eq. \eqref{eq:CS_A} and \eqref{eq:FF_A} respectively.  All known topological electro-magnetic responses in various dimensions can be described by similar topological effective actions\cite{qi2008B}. 

Our primary interest is the 2D CI for which we will use a continuum massive Dirac Hamiltonian as a model.  To calculate the visco-elastic response we couple the massive Dirac Hamiltonian to  geometric perturbations. 
Because of its spinor nature, the Dirac Hamiltonian does not couple to geometry through the metric tensor, but instead via the orthonormal  triad ${\textbf{e}}^{a}$ and its inverse ${\textbf{e}}_{a}$ (frame field) and the spin connection ${\omega^a}_b$. The latin index $a$ labels the particular vector of the frame which, when expanded in terms of a local coordinate basis $\partial/\partial x^{\mu}=(\partial_t, \partial_x,\partial_y),$ has components $e_{a}^{\mu}.$ In a lattice version of the theory, the frame is defined by the local orbital orientation. The stress response can be thought of as a functional of $e^a$ and ${\omega^a}_b$, but we should not take them to be related to each other as they would be in Riemannian geometry\cite{footnote1}. In particular, we will find that the dissipationless viscosity response of the Dirac model is related to torsion. In the context of  condensed matter 
physics, it is convenient, in fact, to set the spin connection to zero such that the torsion is contained in the properties of the triad alone.

The action and Hamiltonian for continuum 2D massive Dirac fermions coupled to a frame field are
\begin{eqnarray}
S&=&\int d^3 x \det ({\textbf{e}})\bar{\psi}\left(p_{\mu}e^{\mu}_{a}\gamma^{a}-m\right)\psi\nonumber\\
H&=&p_x e^{x}_{a}\Gamma^a+p_y e^{y}_{a}\Gamma^a+m\Gamma^0
\end{eqnarray}\noindent with $a=0,1,2$,  $\gamma^a=(\sigma^z, i\sigma^y,-i\sigma^x)$ and $\Gamma^a=(\sigma^z,\sigma^x,\sigma^y).$ If the frame field is position independent the energy spectrum is simply
$E_{\pm}=\pm\sqrt{p_{1}^2+p_{2}^2+m^2}$
with $p_a=e_{a}^{i}p_{i}.$ This is a gapped insulator when $m\neq 0.$ 
Now we will calculate the off-diagonal response of the stress-energy current  (analogous to $\sigma_{xy}$) due to a perturbation of the triad $e^{a}_{\mu}(x)=\delta^{a}_{\mu}+\delta e^{a}_{\mu}(x)$ around the trivial background.  We will see later that the triad  has a simple interpretation in terms of elasticity theory and provides a natural geometric deformation. The stress-energy \emph{current}  that couples to the triad is $T^{\mu}_{a}=\frac{1}{\det ({\textbf{e}})}\frac{\delta S}{\delta e^{a}_{\mu}}=\bar{\psi}p^{\mu}\gamma_{a}\psi.$ 
We wish to integrate out the massive fermions to get an effective action which is a functional of the triad. We are only interested in the terms which lead to dissipationless transport and we find, at leading order,
 \begin{eqnarray}
 \langle T^{\mu}_{a} T^{\nu}_{b}\rangle (q)=\frac{1}{16\pi}\eta_{ab}\epsilon^{\mu\nu\sigma}q_{\sigma}I_{T}(m)\\
 I_{T}(m)=\int_{0}^{\infty}dy\; y \frac{\partial}{\partial y}\frac{m}{(y+m^2)^{1/2}}
 \end{eqnarray}\noindent where $\eta_{ab}={\textrm{diag}}[1, -1, -1]$ is the flat-space Lorentz metric, $q$ is the external momentum, and $y=\vec p^2$ where $p$ is an internal loop momentum.  If we Fourier transform back to real-space this kernel leads to an effective action
 \begin{eqnarray}
 S_{eff}[e^{a}_{\mu}]=\frac{1}{32\pi}I_{T}(m)\int d^3x\; \epsilon^{\mu\nu\rho} e^{a}_{\mu}\partial_{\nu}e^{b}_{\rho}\eta_{ab}.\label{eq:CS_e}
 \end{eqnarray}\noindent which is similar to Eq. \eqref{eq:CS_A}, \emph{i.e.}, 
 a Chern-Simons (CS) term for the triad.
Restoring the spin connection, the integral 
in Eq.\eqref{eq:CS_e} is the Lorentz invariant integral $\int e^a\wedge T^b\eta_{ab}$, with $T^a$ the torsion 2-form.
  For reasons we will see below, we call this a quantum Hall viscosity response.

 When probed by an electric field the 2D continuum Dirac model is notorious for having a half-integer QHE  ($\sigma_{xy}={\rm{sgn}}(m) e^2/2h$) which is connected to the parity anomaly\cite{redlich1984}. However, when properly regularized, (\emph{e.g.} on a lattice) $\sigma_{xy}$ becomes quantized in integer units, as it must for a non-interacting system\cite{Haldane1988}. In the present case, in the continuum limit, the coefficient  $I_{T}(m)$ is  divergent.  If we simply cut off the momentum integral at  a UV scale $\Lambda$ then we find  $I_{T}(m,\Lambda)=-m\Lambda +2m^2{\textrm{sgn}}\; m +O(1/\Lambda).$ Comparing to the quantized Hall conductance,  this is quite different, although from symmetry and dimensional analysis there is  no choice: this term must break time-reversal and thus is an odd function of $m.$ Additionally since $e_{\mu}^{a}$ is dimensionless (unlike $A_\mu$) this coefficient must have units of $1/[{\textrm{Length}}]^2.$ Hence the leading term is proportional to $m$ and the only other scale $\Lambda.$ The other unusual thing is that this term is continuous at $m=0,$ unlike the Hall-conductance, which jumps. To get physically sensible answers for the Hall viscosity, and the Hall conductance, which cannot be half-integer, we must more carefully regulate the theory. Here, we describe the standard Pauli-Villars technique with a set of $N$ massive regulator fields, which is appropriate since it preserves all the symmetries of the Hamiltonian.  The $i$-th  regulator field has mass $M_i$ and wave-function renormalization $C_i$ and we take $M_0=m,\; C_0=1.$ The regulated Hall conductance and viscosity are 
 \begin{equation}
 \sigma_{xy}=\frac{e^2}{2h} \sum_{i=0}^{N} C_i\; {\rm{sgn}}\; M_i, \;\;\;\;\zeta_{reg}=\frac{1}{16\pi} \sum_{i=0}^{N}C_i I_{T}(M_i)\end{equation}\noindent respectively. We can rewrite
 \begin{equation}
 I_{T}(M)=-\lim_{\epsilon\to 0}\frac{M}{\sqrt{\pi}}\int_{\epsilon}^{\infty}dt\; t^{1/2}\int_{0}^{\infty}dy\; ye^{-t(y+M^2)}
 \end{equation}\noindent which yields $I_{T}(M)=-2M/\sqrt{\pi\epsilon} +2 \vert M\vert^2 {\rm{sgn}} M +O(\sqrt{\epsilon}).$ We have three physical constraints that will give proper renormalization conditions (i) $\sigma_{xy}=ne^2/h$, $n\in\mathbb{Z}$ (ii) $\zeta_{reg}$ is finite and (iii) if $\sigma_{xy}=0$ then $\zeta_{reg}=0.$ 
 We know that when the Dirac mass switches sign we go through a phase transition between a trivial insulator and a topological insulator. The sign of $m$ that gives the topological insulator is regularization dependent, and without loss of generality we pick $m>0$ to be the non-trivial insulator. This means that for $m<0$ we require both $\sigma_{xy}=\zeta_{reg}=0$, which is the origin of the third condition. A solution for $M_i, C_i$ under the constraints above is always possible. In all cases, one finds that in the non-trivial phase we  have \begin{equation}
 \sigma_{xy}=\frac{e^2}{h},\;\;\;\;\; \zeta_{reg}=\frac{\hbar}{8\pi} \left(\frac{\vert m\vert}{\hbar v_{F}}\right)^2\equiv\frac{\hbar}{8\pi \ell^2}.
 \end{equation}\noindent If we had chosen $m<0$ to be the topological phase then the signs of both coefficients would be flipped. Note that we have restored the units in the viscosity response. The dimensions of this coefficient are angular momentum density, which is equivalent to momentum per unit length, and the units of dynamic viscosity (force/velocity).  Comparing to the value of the Hall viscosity for the IQHE\cite{avron1995,read2009} $\zeta_{IQHE}=\frac{\hbar}{8\pi\ell_{B}^2}$ we see a similar structure coming from the length scale endowed by the time-reversal breaking field. Also, the viscosity is continuous in the limit $m\to 0$ analogous to the $B\to 0\; (\ell_{B}\to \infty)$ limit of the IQHE.

To make contact with the 
literature on the IQH viscosity we will re-derive the Hall viscosity for the CI state using an adiabatic transport calculation. This can be carried out by putting the Dirac equation on a torus and calculating the adiabatic curvature due to shear deformations   (equivalently deformations of the modular parameter $\tau$) of the torus\cite{avron1995}. 
Consider a square torus, made in $\mathbb{R}^2$ by identifications $(x,y)\sim (x+a,y+b)$ with $a,b\in\mathbb{Z}$, 
with fixed unit volume, and consider area preserving diffeomorphisms, 
corresponding to spatial metrics of the form
\begin{equation}
g_{ij}=\frac{1}{\tau_2}\begin{pmatrix}1&\tau_1\cr\tau_1 &|\tau|^2\end{pmatrix},\ \ \ \ \ 
g^{ij}=\begin{pmatrix}\frac{|\tau|^2}{\tau_2}&-\frac{\tau_1}{\tau_2}\cr-\frac{\tau_1}{\tau_2} &\frac{1}{\tau_2}\end{pmatrix}
\end{equation}
The basis vectors and the spatial part of the triad are 
\begin{eqnarray}
e_1&=&\sqrt{\tau_2}\partial_{x},\ \ \ \ \ 
e_2=\frac{1}{\sqrt{\tau_2}}\left(-\tau_1\partial_{x}+\partial_{y}\right)\\
e^1&=&\frac{1}{\sqrt{\tau_2}} (dx-\tau_1dy),\ \ \ \ \ 
e^2=\sqrt{\tau_2}dy
\end{eqnarray}\noindent respectively, and 
 the Hamiltonian is 
\begin{equation}
H= \begin{pmatrix} m &{\cal{P} }\cr 
{\cal{\bar P}}& -m\end{pmatrix}
\end{equation}
where ${\cal P}=\frac{1}{\sqrt{\tau_2}}\left(\bar\tau p_1-p_2\right).$ We define
${\cal P\cal\bar P}\equiv ||{\cal{P}}||^2.$

We consider the ground state in which all of the negative energy states $\psi_-(p_1,p_2;\tau)$ are occupied. 
The adiabatic connection can be calculated from the explicit form of the single-particle wavefunctions and we find
\begin{eqnarray}
A&=& i\sum_{m,n\in\mathbb{Z}}\psi^\dagger_-(m,n;\tau)d\psi_-(m,n;\tau)\nonumber\\
&=&-i\sum_{m,n\in\mathbb{Z}} f(||{\cal{P}}||^2) \frac12d\ln\left(\frac{{\cal P}}{{\cal\bar P}}\right)
\end{eqnarray}
where
\begin{equation}
f(||{\cal{P}}||^2)=\frac{m}{(m^2+||{\cal{P}}||^2)^{1/2}}
\end{equation}\noindent and where the sums are over the discrete quantized momenta on the torus.
This gives the adiabatic curvature
 \begin{eqnarray}
 F&=&i\frac{d\tau\wedge d\bar\tau}{2\tau_2}\sum_{m,n}p_1^2 f'(||{\cal{P}}||^2).
 \end{eqnarray}
If we convert the sum into an integral we find
\begin{eqnarray}
 F&=&i\frac{d\tau\wedge d\bar\tau}{\tau^{2}_2}\frac{I_{T}(m)}{16\pi}
 \end{eqnarray}\noindent which yields the same (unregulated) viscosity as above.

We will now give a physical interpretation 
in terms of conventional elasticity fields\cite{landauElasticity}.  Assuming we have an elastic medium, we can pick a reference un-displaced state and define a (space-time) displacement field $u^{a}(x).$ Then the triad can be written as $e^{a}_{\mu}=\delta^{a}_{\mu}+w^{a}_{\mu}$ where $w^{a}_{\mu}=\partial u^a/\partial x_\mu$ is the \emph{distortion tensor}\cite{landauElasticity}. To simplify 
we assume that $u^{0}\equiv 0$ \emph{i.e.} $e^{a}=dt.$ Now 
$w_\mu^a$ contains the velocity field $w^{A}_{0}=v^{A}$ and the spatial distortion tensor $w^{A}_{i}$ where $A=1,2$ and $i=x,y.$ This formulation of the triad in terms of the  distortion tensor is consistent with the usual definition as can be seen by calculating the spatial metric $g_{ij}=e^{A}_{i}e^{B}_{j}\delta_{AB}=\delta_{ij}+w_{ij}+w_{ji}+w_{i}^{A}w_{j}^{B}\delta_{AB}$ which 
matches the metric 
from elasticity theory\cite{landauElasticity}.  The stress-energy response from Eq. \eqref{eq:CS_e} is
\begin{eqnarray}
\langle T^{\mu}_{a}\rangle=\zeta_{reg}\eta_{ab}\epsilon^{\mu\nu\rho}\partial_{\nu}e^{b}_{\rho}.
\end{eqnarray}\noindent Since $e^{0}_{\mu}$ does not enter, this simplifies to a momentum-density $\langle T^{0}_{A}\rangle=\zeta_{reg}\eta_{AB}\epsilon^{ij}\partial_{i}e^{B}_{j}$  and a momentum-current $\langle T^{i}_{A}\rangle=\zeta_{reg}\eta_{AB}\epsilon^{i\nu\rho}\partial_{\nu}e^{B}_{\rho}$. These satisfy the continuity equations $\partial_t \langle T^{0}_{A}\rangle=-\partial_i \langle T_{A}^{i}\rangle .$ Restricting ourselves to linear elasticity theory we can freely switch between frame ($a$) and local coordinate ($
\mu$) indices in the response equations. 
Also, space-time indices are raised/lowered using the unperturbed metric.  

  \begin{figure}[t!]
\includegraphics[width=0.4\textwidth]{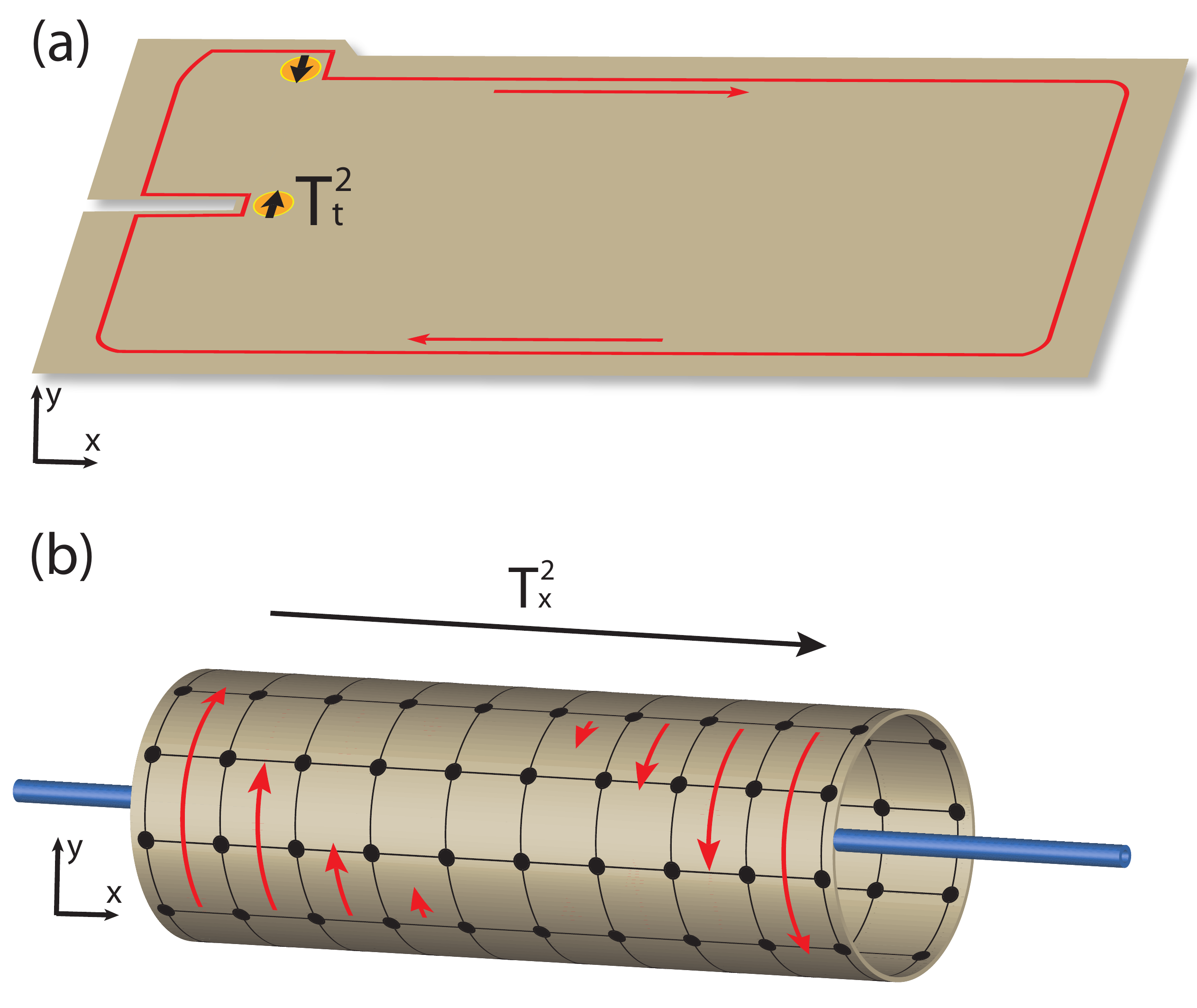}\\
\caption{(a) Chern insulator deformed by a dislocation -antidislocation pair, separated in the $y$-direction. 
For each dislocation, the momentum-density is in the direction of the 
Burgers vector. (b) Chern insulator on a cylinder with a (non-uniform) dislocation threading the cylinder. Local displacements are shown by red arrows. This gives rise to a momentum-current response along the cylinder which carries a momentum component parallel to the Burgers vector of the threaded dislocation, \emph{i.e.}, parallel to the red arrows.}\label{fig:dislocation}
\end{figure}

For $\langle T^{0}_a\rangle\neq 0$,
 $u^A$ 
 cannot be single-valued: it is
a dislocation with Burgers vector $b^{A}$ 
at ${\textbf{x}}_0,$ for which $\epsilon^{ij}\partial_{i}w_{j}^{A}=\epsilon^{ij}\partial_{i}e_{j}^{A}=-b^{A}\delta^{(2)}({\textbf{x}}-{\textbf{x}}_0)$. Thus, the momentum-density response simplifies to $\langle T_{0j}\rangle=-\zeta_{reg}\sum_{m}b^{(m)}_{j}\delta^{(2)}({\textbf{x}}-{\textbf{x}}_m)$ where the ${\textbf{x}}_m$ are the locations of dislocations and $b^{(m)}_{j}$ is the Burgers vector 
of the $m$-th dislocation.  
For a lattice system, the dislocation 
is the fundamental quantized unit of torsion since transporting an electron around a defect 
translates the wavefunction by a multiple of a lattice constant. An edge dislocation with  $\vert {\textbf{b}}\vert=d_b$
contains a momentum density of $\frac{\hbar}{8\pi\ell}\frac{d_b}{\ell}$ along the direction of ${\textbf{b}}.$  
As illustrated in Fig. \ref{fig:dislocation}(a), 
this response is a \emph{momentum} density bound to a torsion `flux' analogous to \emph{charge} density bound to an electromagnetic flux in the bulk of a CI.  Note that Fig.\ref{fig:dislocation} is heuristic, since a realistic edge dislocation is not simply a cut into the material. 

The physical interpretation of the momentum current response $ \langle T_{ij}\rangle$ is not as simple because it is more difficult to 
picture a torsion electric field. In the 2D plane, a moving dislocation (torsion flux) will generate a torsion electric field via the analog Faraday effect. Since we have seen that dislocations naturally carry a momentum density, moving it will  generate a momentum-\emph{current} density as per the response equation. In fact, the momentum-current due to the moving dislocation is being carried perpendicular to the induced torsion electric field.  

Another realization of the momentum current response is obtained by using another instance of the Faraday effect: roll the CI into a cylinder and then insert a torsion flux into the cylindrical hole. This can be thought of as threading a dislocation into the hole of the cylinder so that any particles traveling around the hole will be translated by the Burgers vector $b_a$ of the threaded dislocation. Changing $b_a$ as a function of time creates a torsion electric field the same way that a changing magnetic flux causes a circulating electric field. There is one key difference with the electromagnetic case: in order to preserve area the threaded torsion flux cannot be uniform and the translation must average to zero over the length of the cylinder as illustrated in Fig. \ref{fig:dislocation}(b). Thus, the natural experiment is a torque experiment where a cylinder of CI is twisted. This is equivalent to threading a dislocation with a position dependent Burgers vector.

The formalism developed here 
is a natural generalization of classical elasticity theory. If 
$w_\mu^a =0$ on the boundary,  we can rewrite the effective Lagrangian (Eq. \eqref{eq:CS_e}) as 
 \begin{eqnarray}
 &&\mathcal{L}_{eff}= \frac{\zeta_{reg}}{2}
 \epsilon^{\mu\nu\rho}\eta_{ab}w_{\mu}^{a}\partial_{\nu}w_{\rho}^{b}\nonumber\\ &&=\frac{\zeta_{reg}}{2}
 \epsilon^{\mu\nu\rho}\eta^{\sigma\lambda}\left(u_{\mu\sigma}\partial_{\nu}u_{\rho\lambda}
 + M_{\mu\sigma}\partial_{\nu}M_{\rho\lambda}
 +2M_{\mu\sigma}\partial_{\nu}u_{\rho\lambda}\right)\nonumber\\
 &&u_{\mu\nu}=\frac{1}{2}\left(\frac{\partial u_{\mu}}{\partial x_{\nu}}+\frac{\partial u_{\nu}}{\partial x_{\mu}}\right),\;
 M_{\mu\nu}=\frac{1}{2}\left(\frac{\partial u_{\mu}}{\partial x_{\nu}}-\frac{\partial u_{\nu}}{\partial x_{\mu}}\right)\nonumber
 \end{eqnarray}
 within linear elasticity theory; $u_{\mu\nu}$ and $M_{\mu\nu}$ are the strain and rotation tensors respectively. The first term is the torsional viscosity which is the Lorentz invariant version of the QH viscosity\cite{avron1995}. The stress-energy tensor response $\langle T_{\mu\nu}\rangle$ is \emph{not} necessarily symmetric and thus does not fit in  
 classical elasticity (which would be independent of $M_{\mu\nu}$). It is natural
 to interpret the stress-response within micropolar (
 Cosserat) elasticity theory which takes the local rotational degrees of freedom of the medium into account\cite{eringen1967,Hehl2007}. The distinction here is clear since
 Dirac fermions couple directly to the triad and not the metric, and the spinor nature of the Dirac equation gives local rotational degrees of freedom to which the triad couples.

Finally, we briefly mention two interesting 3D generalizations, the details of which will be presented elsewhere. The first is an anisotropic extension to 3D with the form
\begin{equation}
S_{eff}[e^a_{\mu}]=\frac{\zeta_{\mu}}{2}\int d^4 x \epsilon^{\mu\nu\rho\sigma}e_{\nu}^{a}\partial_{\rho}e_{\sigma}^{b}\eta_{ab}
\label{eq:3dqhe}\end{equation}\noindent where $\zeta_\mu$ is a vector of viscosity coefficients which is analogous to the 3D IQHE\cite{halperin1987}. IQHE or QAHE states which are `stacked' along a direction perpendicular to the vector $\zeta_{\mu}$ exhibit the viscosity response in Eq. \eqref{eq:3dqhe}. This action is basically equivalent to the 3D viscosity response of Ref. \cite{read2010}. 
For a 3D strong TI we find
\begin{equation}
S_{eff}[e^{a}_{\mu}]=\frac{1}{2}\int d^4 x\;  \zeta^{(3)} \epsilon^{\mu\nu\rho\sigma}\partial_\mu e_\nu^{a}\partial_\rho e_{\sigma}^b\eta_{ab}.
\end{equation}
which is a total derivative unless $\zeta^{(3)}$ is not a constant. Hence, on the surface of a 3DTI (where $\zeta^{(3)}$ has a domain-wall) there will be a dissipationless visco-elastic response. This is expected since the surface also contains a QHE.  In gravity theories with torsion the isotropic term is known as the Nieh-Yan term\cite{nieh1982}  and the anisotropic term is 
the (anisotropic)
extension of the triad CS term. 
We leave  open the question on how to experimentally measure this response in 2DEGs or TIs. Unlike electric charge, momentum is not 
conserved when translation symmetry is broken, as 
in any realistic material. Thus questions about possible quantization or the nature of the topological response are still unanswered although interesting discussions of physical implications have been given\cite{haldane2009A,read2010}.

We thank
J. D. Bjorken, K. Fang, L. Freidel, C. Hoyos-Badajoz, X.-L. Qi, A. Randono, S. Ryu, and D.T. Son for discussions.  This work  was supported in part by NSF grant DMR 0758462 
(TLH,EF),  ICMT (TLH), and  the U.S. Department of Energy 
contract DE-FG02-91-ER40709 (RGL). We thank the Galileo Galilei Institute in Firenze for providing a stimulating environment.

%

\end{document}